\documentclass[a4paper]{jpconf}
\usepackage{graphicx}
\begin{document}
\title{Breakdown of the equivalence between gravitational mass and
energy for a composite quantum body}

\author{Andrei G. Lebed}

\address{Department of Physics, University of Arizona, 1118 E. 4-th Street,
Tucson, AZ 85721, USA} \address{and L.D. Landau Institute for
Theoretical Physics, RAS, 2 Kosygina Street, Moscow, Russia}

\ead{lebed@physics.arizona.edu}

\begin{abstract}
The simplest quantum composite body, a hydrogen atom, is
considered in the presence of a weak external gravitational field.
We define an operator for the passive gravitational mass of the
atom in the post-Newtonian approximation of the general relativity
and show that it does not commute with its energy operator.
Nevertheless, the equivalence between the expectation values of
the mass and energy is shown to survive at a macroscopic level for
stationary quantum states. Breakdown of the equivalence between
passive gravitational mass and energy at a microscopic level for
stationary quantum states can be experimentally detected by
studying unusual electromagnetic radiation, emitted  by the atoms,
supported by and moving in the Earth's gravitational field with
constant velocity, using spacecraft or satellite
\end{abstract}

\section{Introduction}
The notion of gravitational mass for a composite body is known to
be non-trivial in general relativity and related to several
paradoxes. The role of the classical virial theorem in
establishing of the equivalence between averaged over time
gravitational mass and energy is discussed in detail in Refs.[1,2]
for different types of classical composite bodies. In particular,
for electrostatically bound two bodies, it is shown that the
gravitational field is coupled to a combination $3K+2U$, where $K$
is the kinetic energy and $U$ the potential energy. Since the
classical virial theorem states that the following time average is
equal to zero, $\bigl< 2K+U \bigl>_t = 0$, we can conclude that
averaged over time gravitational mass is proportional to the total
amount of energy, $E$ [1,2]:
 \begin{equation}
\bigl< m^g \bigl>_t = \bigl<3K+2U\bigl>_t/c^2 = \bigl<K+U\bigl>_t
/c^2 = E/c^2.
\end{equation}

\section{Goal}
The main goal of our paper is to study the quantum problem
concerning the passive gravitational mass of a composite body. As
the simplest example, we consider a hydrogen atom in the Earth's
gravitational field, where we take into account only kinetic and
Coulomb potential energies of an electron in a curved spacetime.
We claim three main results in the paper (see also Refs. [3,4]).
Our first result is that the equivalence between passive
gravitational mass and energy in the absence of gravitational
field survives, at the macroscopic level, in the quantum case.
More strictly speaking, we show that the expectation value of the
mass is equal to $E/c^2$ for stationary quantum states due to the
quantum virial theorem [5]. Our second result is a breakdown of
the equivalence between passive gravitational mass and energy, at
the microscopic level, for stationary quantum states due to the
fact that passive gravitational mass operator does not commute
with energy operator. As a result, there exist a non-zero
probability that a measurement of passive gravitational mass can
yield a value, which is different from $E/c^2$, given by the
Einstein's equation. Our third result is a suggestion of a
realistic experiment to detect this inequivalence by measurement
of electromagnetic radiation, emitted by a macroscopic ensemble of
hydrogen atoms, supported by and moving in the Earth's
gravitational field, using spacecraft or satellite.

\section{Gravitational mass in classical physics}
Here, we derive the Lagrangian and Hamiltonian of a hydrogen atom
in the Earth's gravitational field, taking into account couplings
of kinetic and potential Coulomb energies of an electron with a
weak gravitational field. Note that we keep only terms of the
order of $1/c^2$ and disregard all tidal effects. Therefore, we
can write the interval, using a weak field approximation [6]:
\begin{equation}
d s^2 = -(1 + 2 \phi/c^2)(cdt)^2 + (1 - 2 \phi/c^2) (dx^2
+dy^2+dz^2), \ \phi = - GM/R ,
\end{equation}
where $G$ is the gravitational constant, $c$ is the velocity of
light, $M$ is the Earth mass, $R$ is the distance between center
of the Earth and center of mass of a hydrogen atom (i.e., a
proton). Then, in the local proper spacetime coordinates, which in
the first approximation in $\phi/c^2$ are
\begin{equation}
x'=(1-\phi/c^2) x, \ y'= (1-\phi/c^2) y, z'=(1-\phi/c^2) z , \ t'=
(1+\phi/c^2) t,
\end{equation}
the classical Lagrangian and action of an electron have the
following standard forms:
\begin{equation}
L' = -m_e c^2 + \frac{1}{2} m_e ({\bf v'})^2 + \frac{e^2}{r'} \ ,
\ \ \ S' = \int L' dt' ,
\end{equation}
where $m_e$ is the bare electron mass, $e$ and ${\bf v'}$ are the
electron charge and velocity, respectively; $r'$ is a distance
between electron and proton. It is possible to show that the
Lagrangian (4) can be rewritten in coordinates $(x,y,z,t)$ as
\begin{equation}
L = -m_e c^2 +  \frac{1}{2}m_e{\bf v}^2+\frac{e^2}{r} - m_e \phi -
\biggl( 3m_e\frac{{\bf v}^2}{2}-2\frac{e^2}{r} \biggl)
\frac{\phi}{c^2} ,
\end{equation}
which corresponds to the following Hamiltonian:
\begin{equation}
H = m_e c^2 + \frac{{\bf p}^2}{2m_e}-\frac{e^2}{r} + m_e  \phi +
\biggl( 3 \frac{{\bf p}^2}{2 m_e} -2\frac{e^2}{r} \biggl)
\frac{\phi}{c^2}.
\end{equation}

\section{Gravitational mass in quantum physics}
The Hamiltonian (6) can be quantized by substituting a momentum
operator, $\hat{\bf p} = - i \hbar \partial /\partial {\bf r}$,
instead of canonical momentum, ${\bf p}$:
\begin{equation}
\hat H = m_e c^2 + \frac{\hat {\bf p}^2}{2m_e}-\frac{e^2}{r} +
\hat m^g_e \phi \ ,
\end{equation}
where we introduce passive gravitational mass operator of an
electron to be proportional to its weight operator in a weak
gravitational field (2),
\begin{equation}
\hat m^g_e  = m_e + \biggl(\frac{\hat {\bf p}^2}{2m_e}
-\frac{e^2}{r}\biggl)\frac{1}{c^2} + \biggl(2 \frac{\hat {\bf
p}^2}{2m_e}-\frac{e^2}{r} \biggl) \frac{1}{c^2} \ .
\end{equation}
Note that the Hamiltonian (7),(8) contains the virial contribution
(i.e., the last term) to the mass operator and, therefore, does
not commute with electron energy operator, taken in the absence of
the field. Below, we discuss some consequences of Eqs.(7),(8).
Suppose that we have a macroscopic ensemble of hydrogen atoms with
each of them being in a ground state with energy $E_1$. Then the
expectation value of the electron mass operator (8) per atom is
\begin{equation}
<\hat m^g_e> = m_e + \frac{ E_1}{c^2}  + \biggl< 2 \frac{\hat {\bf
p}^2}{2m_e}-\frac{e^2}{r} \biggl> \frac{1}{c^2} = m_e +
\frac{E_1}{c^2}  ,
\end{equation}
where the third term in Eq.(9) is zero in accordance with the
quantum virial theorem [5]. Therefore, we conclude that the
equivalence between passive gravitational mass and energy survives
at a macroscopic level for stationary quantum states. Let us
discuss how Eqs.(7),(8) break the equivalence between passive
gravitational mass and energy at a microscopic level. Here, we
illustrate the above mentioned inequivalence, using the following
thought experiment. Suppose that at $t=0$ we create a ground state
wave function of a hydrogen atom, corresponding to the absence of
gravitational field,
\begin{equation}
\Psi_1(r,t) = \Psi_1(r) \exp(-iE_1t/\hbar) \ .
\end{equation}
In gravitational field (2), wave function (10) is not anymore a
ground state. It is possible to show that a general solution of
the Schr\"{o}dinger equation, corresponding to the Hamiltonian
(7),(8), can be written as
\begin{equation}
\Psi(r,t) = (1-\phi/c^2)^{3/2} \sum^{\infty}_{n = 1} a_n \Psi_n
[(1- \phi/c^2)r] \exp[-i m_e c^2 (1+\phi/c^2) t/\hbar]  \exp[-i
E_n(1+\phi/c^2) t/\hbar] .
\end{equation}
We wish to draw attention to the fact that wave function (11) is a
series of eigenfunctions of passive gravitational mass operator
(8). $\Psi_n(r)$ is a normalized wave function of an electron in a
hydrogen atom in the absence of gravitational field, corresponding
to energy $E_n$.

In accordance with quantum mechanics, probability that, at $t>0$,
an electron occupies excited state with energy $m_e
c^2(1+\phi/c^2) + E_n(1+\phi/c^2)$ is
\begin{eqnarray}
P_n = |a_n|^2, \ a_n = \int \Psi^*_1(r) \Psi_n [(1-\phi/c^2)r] d^3
{\bf r} = - ( \phi/c^2) \int \Psi^*_1(r) r \Psi'_n(r) d^3 {\bf r},
\nonumber\\
\int \Psi^*_1(r) r \Psi'_n(r) d^3 {\bf r} = V_{n,1}/ (\hbar
\omega_{n,1}), \ \hbar \omega_{n,1} = E_n-E_1 , \ n \neq 1 ,
\nonumber\\
V_{n,1}= \int \Psi^*_1(r) \hat V({\bf r}) \Psi_n(r) d^3 {\bf r} ,
\ \ \hat V({\bf r}) = 2 \frac{\hat {\bf p}^2}{2 m_e} -
\frac{e^2}{r}.
\end{eqnarray}
Let us discuss Eq.(12). We stress that it directly
demonstrates that there is a finite probability,
\begin{equation}
P_n = |a_n|^2 = (\phi/c^2)^2 \ [V_{n,1}/(E_n-E_1)]^2 \ , \ n \neq
1,
\end{equation}
that, at $t>0$, an electron occupies n-th ($n \neq 1$) energy
level, which breaks the expected Einstein's equation, $m^g_e=m_e +
E_1/c^2$. In fact, this means that quantum measurement of passive
gravitational mass in a quantum state with a definite energy (10)
gives the following quantized values:
\begin{equation}
m^g_e (n) = m_e + E_n/c^2  \ .
\end{equation}

\section{Suggested experiment}
Here, we describe a realistic experiment: We consider a hydrogen
atom being in its ground state at $t=0$, located at distance $R'$
from the center of the Earth and having the wave function:
\begin{equation}
\tilde{\Psi}_1(r,t) = (1-\phi'/c^2)^{3/2} \Psi_1[(1-\phi'/c^2)r]
\exp[-im_ec^2(1+\phi'/c^2) t /\hbar]
\exp[-iE_1(1+\phi'/c^2)t/\hbar] .
\end{equation}
The atom is supported by in the Earth's gravitational field and
moving from the Earth at constant velocity, $u \ll \alpha c$
(where $\alpha$ is the fine structure constant), by spacecraft or
satellite. It is possible to show that the electron wave function
and the time dependent perturbation for the Hamiltonian (7)-(8),
in the coordinate system related to center of mass of the atom,
can be expressed as
\begin{eqnarray}
\tilde{\Psi}(r,t) = (1-\phi'/c^2)^{3/2} \sum^{\infty}_{n=1}
\tilde{a}_n(t) \Psi_n[(1-\phi'/c^2)r] \exp[-im_ec^2(1+\phi'/c^2) t
/\hbar]
\nonumber\\
\exp[-iE_n(1+\phi'/c^2)t/\hbar] ,
\end{eqnarray}
\begin{equation}
\hat U ({\bf r},t) =\frac{\phi(R'+ut)-\phi(R')}{c^2}  \biggl(3
\frac{\hat {\bf p}^2}{2m_e}-2\frac{e^2}{r} \biggl) .
\end{equation}
We wish to draw attention that in a spacecraft (satellite), which moves
with constant velocity, gravitational force, which acts on each
individual hydrogen atom, is compensated by some non-gravitational
forces. This causes very small changes of a hydrogen atom energy
levels and is not important for our calculations. Therefore, the
atoms do not feel directly gravitational acceleration, ${\bf g}$,
but feel, instead, gravitational potential, $\phi(R")=  \phi
(R'+ut)$.

It is important that, if excited levels of a hydrogen atom were
strictly stationary, then probability to find the passive
gravitational mass to be quantized with $n \neq 1$ (14) would be
\begin{equation}
\tilde{P}_n = (V_{n,1}/\hbar \omega_{n,1})^2
 [\phi(R')/c^2]^2   \ , n \neq 1, \ |\phi(R'+ut)| \ll |\phi(R')|,
 u \ll \omega_{n,1} R ,
\end{equation}
which coincides with Eq. (13). In reality, the excited levels
spontaneously decay with time and, therefore, it is possible to
observe the quantization law (14) indirectly by measuring
electromagnetic radiation from a macroscopic ensemble of the
atoms. In this case, Eq.(18) gives a probability that a hydrogen
atom emits a photon with frequency $\omega_{n,1} = (E_n-E_1) /
\hbar$ during the time of the experiment. To estimate the
probabilities (18) we use the following numerical values of the Earth's
mass, $M \simeq 6 \times 10^{24} kg$, and its radius, $R_0 \simeq
6.36 \times 10^6 m$. It is important that, although the
probabilities (18) are small, the number of photons, $N$, emitted
by macroscopic ensemble of the atoms, can be large. For instance,
for 1000 moles of hydrogen atoms, $N$ is estimated as
\begin{eqnarray}
N (n \rightarrow 1) = 2.95 \times 10^{8} \ [V_{n,1}/(E_n-E_1)]^2 , \
\ N (2 \rightarrow 1) = 0.9 \times 10^8 ,
\end{eqnarray}
which can be experimentally detected. [Here, $N(n \rightarrow 1)$
stands for a number of photons, emitted with frequency $\omega_{n,1} = (E_n -E_1)/\hbar$.]

\ack

We are thankful to N.N. Bagmet (Lebed), V.A. Belinski, Steven Carlip,
Li-Zhi Fang, Douglas Singleton, and V.E. Zakharov for useful
discussions. This work was supported by the NSF under Grant
DMR-1104512.

\end{document}